\documentclass[floatfix,preprintnumbers,superscriptaddress,showpacs,
               showkeys,prl,preprint]{revtex4} 

\usepackage[version=3]{mhchem} 
\usepackage{graphicx}
\usepackage{dcolumn}
\usepackage{amsmath, amsfonts, amssymb, bm}

\begin{document}
\bibliographystyle{apsrev} 


\title{Cubic to hexagonal iron phase transition
promoted by interstitial hydrogen
} 

\author{A. Castedo}
\author{J. Sanchez}

\affiliation{
Instituto de Ciencias de la Construccion Eduardo Torroja,
IETcc-CSIC, Madrid (Spain)
}

\author{J. Fullea}
\author{M.C. Andrade}

\affiliation{
Centro de Seguridad y Durabilidad Estructural y de Materiales,
CISDEM-UPM-CSIC, Madrid (Spain)
}

\author{P.L. de Andres
\footnote{On leave of absence from Instituto de Ciencia de Materiales de Madrid  (CSIC) 28049 Madrid (Spain)}
}

\affiliation{
Donostia International Physics Center (DIPC),
Paseo Manuel Lardizabal 4, 20018 San Sebastian (Spain)
}

\date{\today}

\begin{abstract}
Using ab-initio density functional theory we study the role of interstitial hydrogen on the
energetics of the phase transformation of iron from bcc to hcp along Bain's  pathway.
The impurity creates an internal stress field that can be released through
a tetragonal distortion of the lattice, promoting the 
bcc (ferromagnetic) $\rightarrow$ fcc (frustrated antiferromagnetic) 
$\rightarrow$ hcp (ferromagnetic) transition.  
The transformation between crystal systems is accompanied by a drastic magnetic reorganization and sudden variations of the unit cell volume, 
that can be one of the reasons for embrittlement and mechanical failure
of iron upon hydrogen adsorption.
\end{abstract}

\pacs{62.20.mj,64.70.K-,,66.30.J-,81.30.Kf}




%

\keywords{iron, hydrogen, interstitial,
bcc ($\alpha$-iron), fcc ($\gamma$-iron), hcp ($\epsilon$-iron),
fragilization, embrittlement, ab-initio, density functional, phase transition}


\maketitle

{\it Introduction.}
Interstitial impurities play a key role on mechanical and structural properties of different metals.
In particular, interstitial hydrogen has been identified as a promoter of mechanical failure under stress; a point of particular interest in the technological areas of engineering and 
construction.
These processes, known by the generic name of embrittlement, 
are subject to active research in the field of high-strength steels, 
of which the primary component is ferromagnetic body-centered cubic  iron 
(bcc, also referred as $\alpha$-iron). 
\cite{eliaz02,liang03,Ohno03,elices04,Sanchez07,robertson11}
Furthermore, hydrogen dissolved in iron at very high pressures is fundamental 
to understand the behavior of the Earth's core, 
in particular the long-standing controversy about its 
density and even its crystalline phase.\cite{gillan02,johansson07}
Experimental information in these areas is difficult to obtain and sometimes contradictory. Therefore, a key element to make progress in such complex but fundamental problems is the use of accurate ab-initio density functional theory (DFT), 
that can provide guiding in the interpretation and new motivation 
for gathering further experimental evidence.


In the cubic lattice there are two high-symmetry sites competing to host
interstitial impurities: the tetrahedral and octahedral sites.
Absorption in the octahedral site creates an anisotropic stress field that favors
a tetragonal distortion, making the bcc configuration unstable and driving the unit 
cell towards a phase transition to a face-centered cubic lattice (fcc)
first, and then an hexagonal one (hcp). 
Absorption on the tetrahedral site, on the other hand, results in a more isotropic
stress distribution and merely provokes an increase of the volume.
There has been an ongoing controversy in the literature about the 
actual absorption site for interstitial hydrogen in ferromagnetic bcc iron.
This is partly related to the fact that the two competing
available high-symmetry sites are very similar in energy,
making difficult to accurately decide between them because of the many 
different factors involved. In particular, as the computed energy difference
between them is similar to $k_{B} T$ at room temperature, an 
equilibrium distribution between both sites should exist. 
Furthermore, external stresses can contribute to the energy enough
to influence that distribution, and a many-body contribution depending
on the interstitials density can influence the final distribution.
Currently accepted wisdom takes the tetrahedral site (T) as the most likely 
absorption site for the low-temperature, low-density phase,  while the octahedral 
site (O) is preferred for  large concentrations of 
interstitials.\cite{carter04,sanchez08, sanchez10} 
From a physical point of view, this is related to the internal stress originated 
from the occupation of O-sites, causing a tetragonal distortion of the bcc lattice 
and making unfavorable the occupation of T-sites for large densities of interstitials. 
Generally speaking, overall hydrogen solubility in bcc iron is low, 
favoring the occupancy of T-sites. 
However, occupation of O-sites cannot be neglected under some circumstances,  e.g.:
(i) if there is a local accumulation of interstitial hydrogen, e.g. due to
preferential nucleation near defects,
which would keep solubility low over the whole sample but affect
significantly the local characteristics of the material,\cite{castellote07} 
(ii) large solubilities due to high external pressure, as it is believed to be 
the case in the Earth's core,\cite{gillan02}, and finally
(iii) because an externally-driven tetragonal distortion of the lattice
definitively favors the occupation of O-sites over T-sites.
Furthermore, a tetragonal distortion of the required kind happens naturally along
the pathway introduced by Bain to account for the phase transformation
between bcc and fcc iron.\cite{bain24}
A self-sustained scenario can be realized here: 
the tetragonal distortion behind the phase transition
favors the occupation of O-sites, and in turn,
increasing the population at these sites drives the necessary distortion
for the phase transformation.
In this paper, we shall study the modification of
kinetics barriers for a bcc $\rightarrow$ fcc $\rightarrow$ hcp transformation 
due to the  role of interstitials occupying O-sites.

\begin{figure}
\includegraphics[clip,width=0.99\columnwidth]{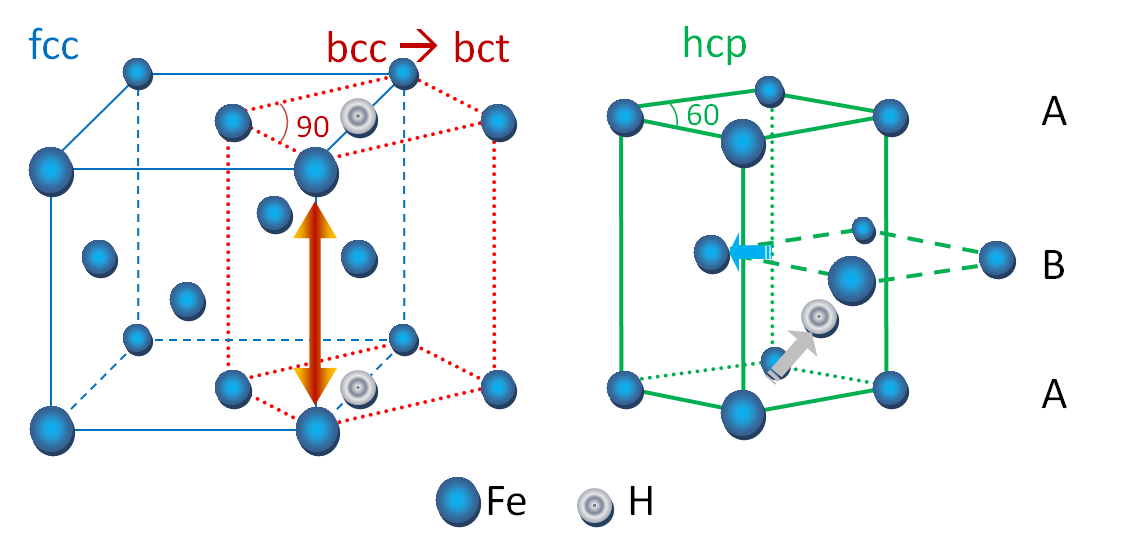}
\caption{(color online) Unit cell sketch along
Bain's transformation:  bcc ($c/a=1$, middle),
fcc ($c/a=\sqrt{2}$, left),  hcp ($c/a=1.6$, right).
The Octahedral high symmetry site (O)
is shown as grey circle. 
}
\label{fgr:Fig1}
\end{figure}

The tetragonal distortion of the bcc lattice belongs to the group
of martensitic transformations, which represent a diffusion-less, 
cooperative, homogeneous movement of the atoms providing a 
change in the crystal structure  that can be monitored by symmetry operations, 
lattice distances and or unit cell volumes.\cite{bain24,sliwko96}
Adsorbing hydrogen in O-sites deforms the bcc lattice into 
a body-centred-tetragonal (bct) phase (Fig. \ref{fgr:Fig1}). \cite{sanchez08}
The tetragonal distortion consists of a contraction upon two of the cubic 
axes ($a=b$), while the third one expands ($c$). 
When the $c/a$ ratio reaches the value $\sqrt{2}$ the bct lattice becomes 
face-centred-cubic (fcc), also known as $\gamma$-Fe, 
a stable frustrated antiferromagnetic phase of iron. 
According to our calculations, 
for values of $c/a \ge 1.5$ and a stoichiometry \ce{Fe2H}
a further transformation to an hexagonal phase 
($\epsilon$-Fe)
takes place
($\gamma = 90^{\circ} \rightarrow 60^{\circ}$;
Fig. \ref{fgr:Fig1}).
Bain's pathway is not the only transformation suggested in the literature 
to explain the phase diagram of iron, 
but it has a number of advantages for the purposes of this work,
namely: 
(i) it retains at each step the highest possible crystal symmetry,
(ii) it is simple, 
and (iii) it contains the lattice deformation 
expected for hydrogen occupation of 
O interstitials.

\begin{figure}
\includegraphics[clip,width=0.99\columnwidth]{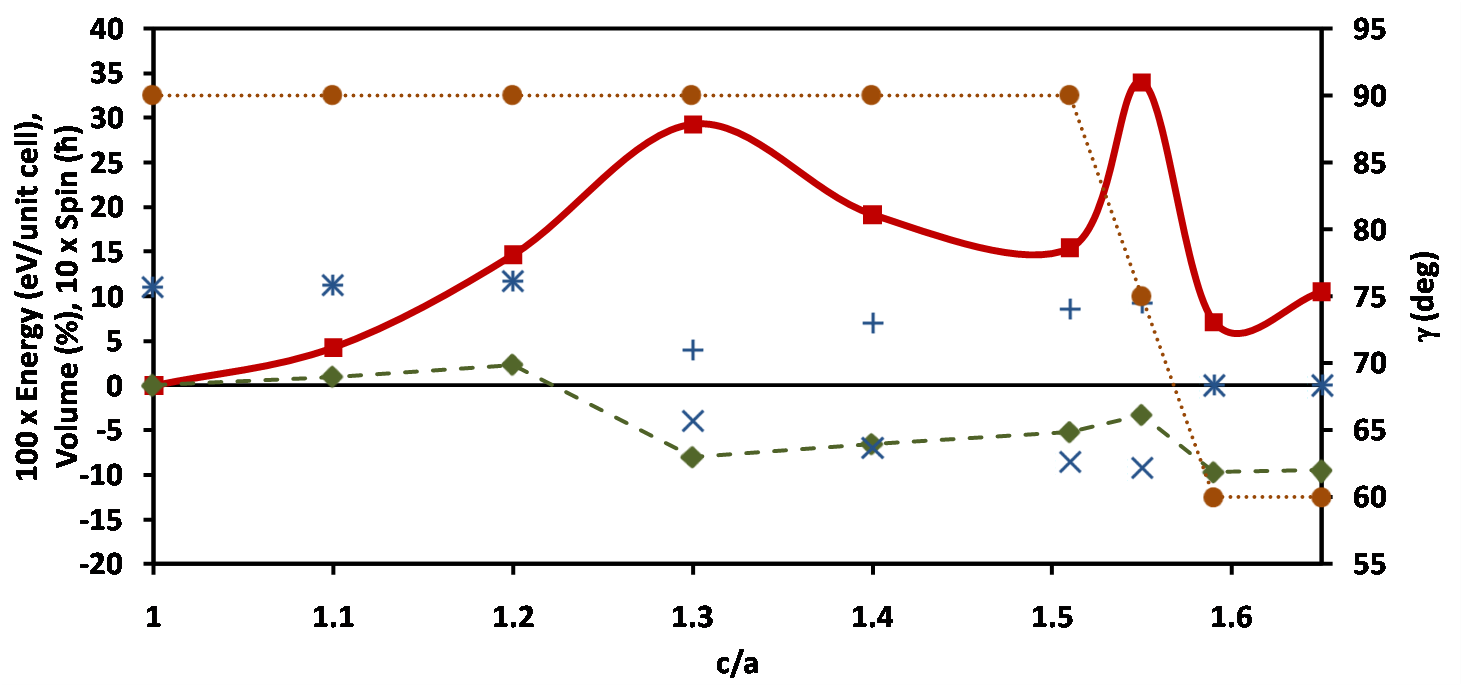}
\caption{(color online)
Along Bain's pathway represented in
Fig. \ref{fgr:Fig1}
(from $c/a=1$ --cubic-- to $c/a=1.6$ --hexagonal--):
(i) $\gamma$ in degrees (right y-axis, brown dotted line and circles),
(ii) enthalpy in eV/unit 
cell (left y-axis, red thick continuos line and squares;
to increase visibility values have been multiplied by 
an arbitrary factor of $100$);
(iii) volume of the unit cell expressed as the fractional
variation with respect to the global minimum value at
$c/a=1$ (right y-axis, green dashed line and diamonds),
(iv) spin population in $\hbar$ on the two iron atoms in the unit cell: 
Fe(1) and Fe(2) (left y-axis, blue $+$ and $\times$  
respectively; to increase visibility the numerical value has 
been multiplied by an arbitrary factor of 10).
Notice how the transformation between ferromagnetic (both spins are positive) 
to antiferromagnetic (spins have opposite signs) to non-magnetic (both are zero)
are related to sudden variations in  the volume of the unit cell.  
Lines joining symbols have been drawn to guide the eye only.
}
\label{fgr:Fig2}
\end{figure}

{\it Theory.}
Ab initio calculations have been carried out in the framework of DFT 
and pseudo-potentials theory. \cite{kohn64,vanderbilt90}
Actual calculations have been performed with the CASTEP code. \cite{clark05}
The Born-Oppenheimer approximation is used; ions are considered classical 
objects moving under the forces created by electrons obeying Schr\"odinger equation. 
Ultrasoft pseudopotentials are used to describe Fe and H,
and wavefunctions are expanded in plane-waves  
with a cut-off energy of 375 eV. 
The generalized gradients approximation for the exchange and correlation 
potential due to Perdew, Burke and Ernzernhof \cite{pbe} has been chosen,
and spin-polarized bands are considered to account for magnetism. 
The accuracy of our calculations is determined by the quality of the pseudo-potential, 
the cut-off energy, and the density of the k-points mesh used in the irreducible part 
of the Brillouin zone (a Monkhorst-Pack mesh of 10x10x10 \cite{monk}). 
The choice of these values for our calculations is related to previous studies 
where conditions for good convergence on relevant properties of iron such 
as the lattice parameter, magnetization, or bulk modulus have been assessed \cite{sanchez08}. 
Other convergence thresholds are: 
(i) variation in the total energy $\le 10^{-5}$ eV/atom, 
(ii) maximum residual force $\le 0.001$ eV/{\AA}, 
(iii) maximum change in any atom position $\le 0.001$ {\AA}. 
All the parameters defining the unit cell (distances and angles) 
have been optimized in order to minimize the stress for each crystal system
(cubic, tetragonal and hexagonal):
local minima have been optimized so the maximal residual pressure on the
unit cell is $P \le 0.01$ GPa
(such a low threshold is convenient to
compute elastic constants around equilibrium
configurations), while barriers have been optimized to a threshold
below $\le 0.1$ GPa, affecting the energy   
$\le 0.007$ eV. 
Barriers between these points have been estimated by a constrained 
optimization procedure where the parameters $a/c$ and $\gamma$
are kept fixed to different values
and all the remaining
parameters (including magnetization) 
are freely optimized so the system can reach
the nearest local minima
(Fig. \ref{fgr:Fig2}) . 

{\it Results.} 
To understand the role of interstitial hydrogen 
we compare the energy landscape of Fe and \ce{Fe2H} for various
crystal symmetries.
At T=0 K and P=0 GPa, the global minimum for Fe corresponds to
ferromagnetic (FM)  bcc Fe ($c/a=1$, $\gamma=90^{\circ}$).
Varying continuously the $c/a$ ratio we find two metastable  
configurations, first  for Fe fct ($c/a \approx 1.51$),
and then for hcp ($c/a \ge 1.6$, $\gamma=60^{\circ}$).
Their magnetic configurations are respectively 
anti-ferromagnetic type I (AF) and non-magnetic (NM)
(Fig. \ref{fgr:Fig2}). 
We have computed structural parameters and
elastic constants for these three
configurations (upper sections in Tables \ref{tab:table1} and  \ref{tab:table3}): 
there is an overall good agreement with the available experimental data
and all the elastic constants are positive and
consistent with the symmetry, proving that they are truly locally
stable minima. 
We notice, however, that theory predicts
a frustrated antiferromagnetic fct phase ($c/a \approx 1.5$),
not a pure fcc cristal ($c/a=\sqrt{2}$) as it has been experimentally
reported.\cite{Wassermann94}
In agreement with other authors
we remark the importance of magnetism to understand
the phase diagram of iron.\cite{hasegawa83,krasko89,Watanabe05,okatov09} 
The contribution to the total energy of magnetism is itself small, 
but crucial since it is similar to differences between local minima and 
barriers ($\le 0.2$ eV/unit cell).
Furthermore, every important change in the magnetic
state (from FM to AF, to NM) is accompanied by a corresponding sudden
change in the unit cell volume: a decrease in the volume increases
the overlap of electrons in the sp-like bands increasing the delocalization
energy (kinetic) and reducing the strength of magnetism
(Fig. \ref{fgr:Fig2}a).

\begin{figure}
\includegraphics[clip,width=0.99\columnwidth]{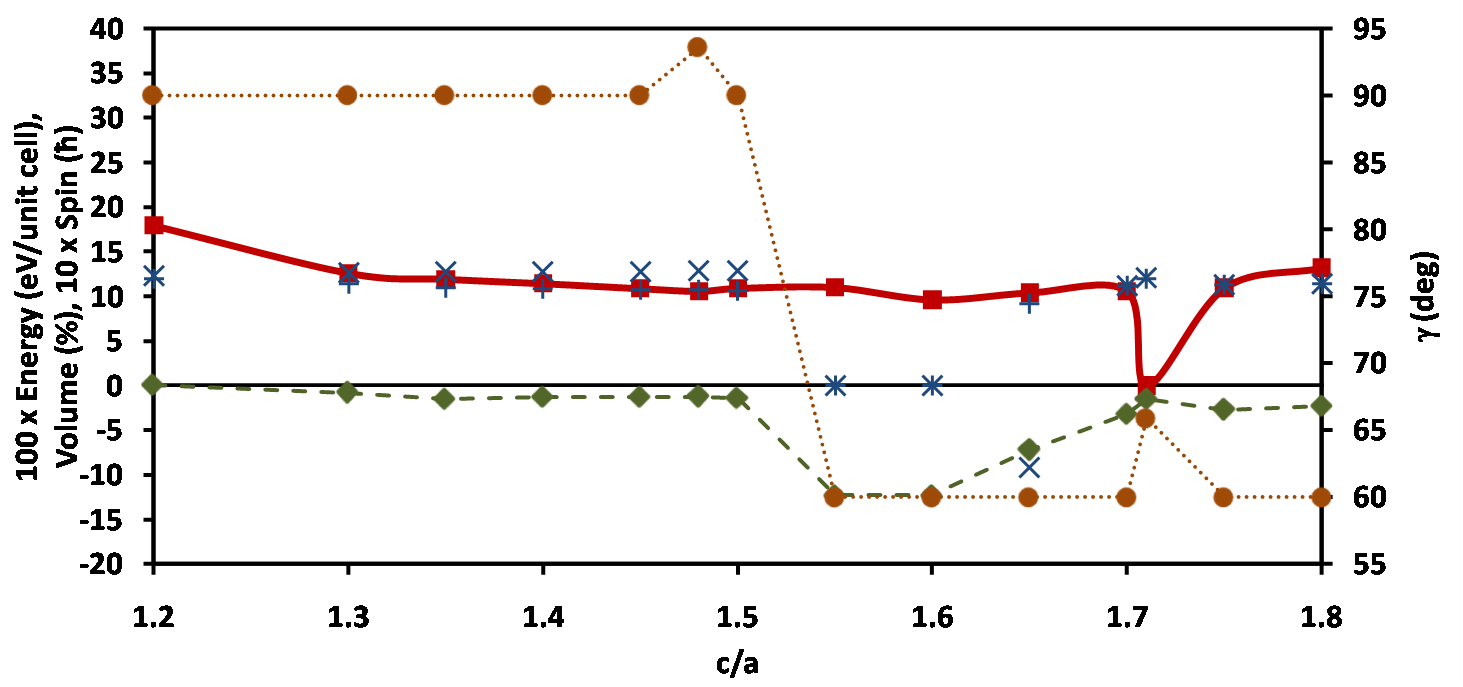}
\caption{(color online)
For \ce{Fe2H}, same as caption in Fig. \ref{fgr:Fig2}.
}
\label{fgr:Fig3}
\end{figure}

Interstitial hydrogen absorbed on the octahedral site 
has an important effect on the above picture.
The internal pressure on the bcc lattice
is approximately given by the stress tensor:
$$
\sigma =
\begin{bmatrix}
10 & 0 & 0 \\
0 & 10 & 0 \\
0 & 0 & -12 \\
\end{bmatrix}
(GPa)
$$
\noindent
It is obvious how this internal stress is
responsible for the bct deformation of 
the unit cell that drives the system along the Bain's pathway
towards the fcc first, and subsequently to
an hexagonal phase reached by a mere shear and a shuffle
of the atoms in the unit cell.\cite{Caspersen04}
We notice in passing that hydrostatic pressures between 10
and 20 GPa mark the onset of the transformation between
the bcc and the hcp phases at low temperatures.\cite{jephcoat86}
The internal pressure derived from the hydrogen
presence is in the same range of values, therefore
the behavior in Fig. \ref{fgr:Fig3},
where barriers for the phase transformation have disappeared,
is not at all surprising.
During this transformation
the magnetic interaction gets weaker in the region
$c/a \approx 1.5 - 1.6$, and the volume
of the unit cell is accordingly reduced by $\approx 15$\%.
This is in good agreement with the experimental evidence
for the $\epsilon$-Fe phase, where it is known that
hydrogen does not absorb in tetrahedral sites,
and magnetism is weak.\cite{antonov98}
In particular, the non-magnetic hexagonal structure
for clean Fe ($c/a=1.63$) 
is believed to exist at high pressure
in the Earth's core.\cite{antonov98}
However, in the presence of interstitials we observe
how the system is pushed further along the
path to reach a sharp minimum at
$c/a = 1.71$ ($\gamma=60^{\circ}$), 
where the volume increases again and the
the ferromagnetism is restored in
the material.
Variations in the volume of the unit cell 
reflect the need of the system
to minimize internal
stresses, but these changes, now nearly twice
the ones for clean iron, facilitate
the appearance of defects in the material.
Similarly to pristine iron, experimental and theoretical
parameters in Tables \ref{tab:table1} and  \ref{tab:table3}
agree well, 
except for the predicted ferromagnetism
for \ce{Fe2H}, that Antonov et al. did not find on 
the hexagonal phase of \ce{Fe2D}. \cite{antonov98}



{\it Conclusions.}
Calculations for the Bain path of clean iron have allowed us to investigate the different 
stable/metastable phases of iron according to the energetic of the structures.  
Interstitial hydrogen hosted on O-sites becomes a source of large internal stress 
that can be released by a body-centered tetragonal deformation 
that fits naturally with Bain's mechanism. 
DFT calculations show that the bcc phase is deformed until it reaches an AF-type I 
phase with $c/a=1.5$ (close to the fcc crystal structure).
It is favorable for the system to continue deforming to reach a paramagnetic minimum
with hexagonal symmetry ($c/a=1.6$ and $\gamma=60^{\circ}$).
This scenario agrees with current ideas about the
Earth's core composition under extreme external pressure.
The internal pressure introduced by interstitial hydrogen can be thought as an
effective temperature given to the system, and can be useful to understand the
iron phase diagram.
It is known that the $\gamma$ to $\alpha$ transformation happens fast up to 
a critical temperature, $M_{s}$, but for higher temperatures it is slowed by
competition with a different process. 
Our calculations show that such a martensitic transformation should compete
with the formation of an $\epsilon$ phase favored by the 
accumulation of interstitials in octahedral sites. 
This process inhibits the $\gamma$ to $\alpha$
phase transition because small nucleation seeds are formed that will grow 
by attracting further hydrogen
diffusing fast at high temperatures.

{\it Acknowledgments.}
This work has been financed by the Spanish
MICINN (MAT2008-1497 and BIA2010-18863), 
and MEC (CSD2007-41 "NANOSELECT"
and "SEDUREC").


\newpage

\clearpage
\newpage

\clearpage
\newpage

\clearpage
\newpage

\begin{table}
\caption{For the different phases of clean iron (above
and \ce{Fe2H} (below)
theoretical and experimental
parameters:
IT is the number of the symmetry group
in the International Tables for Crystallography,
lattice parameter $a$ is given in {\AA} and the unit cell volume in  {\AA}$^{3}$,
enthalpies in eV,
forces in eV/atom, spin in $\mu_{B}$,
and Bulk modulus in GPa.
Notice that for \ce{Fe2H}, the FM and NM hcp are quasi-degenerated
in energy, with FM the slightly more stable configuration at T=0 K.
The fct-AF$^{*}$ phase corresponds to a tetragonal distortion of the fcc
($c/a=1.5$) with a frustrated anti-ferro configuration.
\label{tab:table1}
}
\begin{tabular}{llcrcrc}
\hline
SYSTEM & IT & a ( {\AA})  &   c/a  & s ($\mu_{B}$) &  vol {\AA}$^{3}$& B (GPa) \\
 \hline
 Fe &  &    &     &   &   &   \\
  \hline
BCC                   &  229       & 2.82 & 1.00   & 2.2  & 22.33        & 210 \\
BCC\cite{kittel}  &  229     & 2.87 & 1.00  & 2.2  &  22.80       & 172 \\
FCT            & 139    & 2.43 & 1.50   & 0.0  & 20.27        & 207 \\
FCC\cite{Stassis87} & 225   & 2.55 & 1.42    & 0.0     & 23.55 & 133 \\
HCP                     & 194  & 2.45 & 1.58   & 0.0 & 22.16         & 311  \\
HCP\cite{jephcoat86} & 194                                  & 2.58 & 1.62   & 0.0 &    24.09 & 180  \\
\hline
\ce{Fe2H} &  &      &   &   &   &   \\
\hline
MONOCLINIC                 &  12  & 2.53  & 1.71   & 2.39             & 24.85 & 183  \\
HCP\cite{antonov98}  & 194    & 2.58    & 1.62   & 0.0   & 
24.09  &   -                   \\
\hline\hline
\end{tabular}
\end{table}

\clearpage
\newpage

\begin{table}
\caption{Elastic constants (GPa) computed at the relevant local minima for Fe
and \ce{Fe2H} (upper and lower, respectively).
\label{tab:table3}
}
\begin{tabular}{lrrrrrrrr}
\hline
              & $C_{11}$ & $C_{12}$ & $C_{13}$ & $C_{33}$ &$C_{44}$  &  $C_{55}$&  $C_{66}$  \\
\hline
 Fe & & & & &  & \\
\hline
 TH bcc FM         & 305& 162&   &   & 121   & &  \\
 EXP bcc FM\cite{Singh98} & 281 & 144 &  & & 123  & & \\
 TH fct AF$^{*}$  & 579 & 120   & 265 & 10  & 196 &   & 50   \\
 EXP fcc\cite{Stassis87} & 154 & 122 &  &  & 77  & & \\
 TH hcp NM        & 570& 187& 158    & 661   & 184 &   & 191  \\
 EXP hcp NM\cite{Singh98}  & 640 & 300& 254   & 648   & 422 & &   \\
 \hline
  \hline
\ce{Fe2H} & & & &  & & \\
\hline
 monoclinic, FM & 322 & 124 & 97 &  329  & 104 & 73 & 104      \\
 \hline
 \end{tabular}
 \end{table}

\end{document}